\documentclass[aps,prd,preprint,groupedaddress]{revtex4}
\usepackage{axodraw}
\usepackage{ulem}
\usepackage{slashed}
\usepackage{latexsym}
\usepackage{amssymb}
\usepackage{graphicx}
\usepackage{bm}

\newcommand\beq{\begin{equation}}
\newcommand\eeq{\end{equation}}

\newcommand\half{\frac{1}{2}}

\begin{document}

\preprint{}

\title{Toward a Minimum Branching Fraction for
Dark Matter Annihilation into Electromagnetic Final States}

\author{James B.\ Dent}
\affiliation{Department of Physics and Astronomy,
Vanderbilt University, Nashville, TN 37235, USA}

\author{Robert J.\ Scherrer}
\affiliation{Department of Physics and Astronomy,
Vanderbilt University, Nashville, TN 37235, USA}

\author{Thomas J.\ Weiler}
\affiliation{Department of Physics and Astronomy,
Vanderbilt University, Nashville, TN 37235, USA}

\date{\today}

\begin{abstract}
Observational limits on the high-energy neutrino background have
been used to place general constraints on dark matter that
annihilates only into standard model particles.
Dark matter particles that annihilate into neutrinos
will also inevitably
branch into electromagnetic final states through higher-order tree and loop diagrams
that give rise to charged leptons, and
these charged particles can transfer their energy into
photons via synchrotron radiation or inverse Compton scattering.
In the context of effective field theory, we calculate
the loop-induced branching ratio to charged leptons 
and show that it is
generally quite large, typically $\agt 1$\%,
when the scale of the dark matter mass exceeds the electroweak scale, $M_W$.
For a branching fraction $\agt 3$\%, the synchrotron radiation bounds on dark
matter annihilation are currently stronger than the
corresponding neutrino bounds in the interesting mass range
from 100 GeV to 1 TeV.  For dark matter masses below $M_W$, our
work provides a plausible framework for the construction of a model
for ``neutrinos only" dark matter
annihilations.

\end{abstract}



\maketitle


While dark matter (DM) accounts for 25\% of the total energy density in the universe 
(a classic review is~\cite{jungman}), direct detection remains elusive 
(see, e.g., Ref.~\cite{hooper} for a recent discussion).  
If the dark matter is a thermal relic, one would expect a non-neglible annihilation
cross section, and one might hope to either detect the products of such annihilations
occurring today, or to bound the properties of such particles by
the non-observation of annihilation products.

Recently, Beacom, Bell and Mack~\cite {beacom}
made an interesting argument regarding such limits.
They noted that if a dark matter particle $(\chi)$ couples only to Standard
Model (SM) particles, then the most general limits can be derived by
assuming annihilation into neutrinos, as these are the most difficult
SM annihilation products to detect.  Rather surprisingly, current
observations place nontrivial bounds on such annihilations (see
also Refs.~\cite{yuksel,PR}).  Mack et al.~\cite{mack} compared these
neutrino annihilation limits with corresponding limits on gamma-ray-producing
annihilations.
Their results indicate that, for an illustrative branching ratio to photons of $10^{-4}$,
the neutrino bounds generally provide tighter
constraints for large dark matter masses ($M_\chi >100$ MeV), 
while the photon bounds dominate at smaller masses.
If the branching ratio into photons
is taken to be larger, then
the mass range over which the neutrino bounds dominate
shrinks accordingly.
Further, with the GLAST satellite~\cite{GLAST} expected to launch in 2008, the photon limits
will shortly be considerably tightened (or dark matter annihilation
will be detected!).

Thus, it is very interesting to determine
what constitutes a ``reasonable" minimum branching
ratio into electrically-charged particles, for these in turn generate photons.  
We will assume a
``neutrinos only" final state at tree level, and then calculate the branching
into charged leptons that results from a higher-order box diagram.  Note that a 
calculation similar in spirit was undertaken by Kachelriess and Serpico~\cite{WZstrahlung}, who examined 
the electromagnetic mode from electroweak bremsstrahlung of real $W$ and $Z$ particles.  
In quantitative detail~\cite{bell}, it is found that  
the branching fraction of this process is
$\frac{\alpha}{12\pi} \frac{M_\chi^2}{M_W^2}=2.1\times 10^{-4}\times\frac{M_\chi^2}{M_W^2}$,
for $M_\chi\agt M_W$ such that the electroweak bosons are produced on-shell;
we have taken $\alpha=1/128$
for the numerical value, as is appropriate at the weak scale.
The processes we discuss here
are physically distinct from brehmsstrahlung 
and offer a potentially more significant contribution
to the electromagnetic branching ratio. In this work, 
we obtain a rate ratio from loop graphs growing as 
$\left(\frac{\alpha\,M_\chi^2}{8\pi\,M_W^2}\right)^2$ times a logarithmic factor.

It is non-trivial to embed a tree-level ``neutrinos only'' final state model into a field theory.
The special status afforded the neutrino breaks the $SU(2)$ invariance of the weak interaction.
As a consequence, the embedded theory we construct must be viewed as incomplete, as an 
effective theory at best, valid up to some energy scale $\Lambda$.
The effective theory will also be non-renormalizable, and therefore 
unstable against radiative corrections.  This presents another indication that 
the neutrinos-only model exists only for a finely-tuned effective theory below 
a scale $\Lambda$, 
or for a UV-completed theory with a rich spectrum at $\Lambda \sim M_B$,  
where $M_B$ is the mass scale of the interaction connecting the DM sector to neutrinos.
The rich-spectrum possibility seems contrived and unlikely, but if true, it predicts 
much discovery not far beyond the energy reach of today's accelerators.
We will assume that the neutrinos-only model is simple  
in a region of validity up to $\Lambda^2\gg M^2_B$.  
We will present the consequences of such a model.

Assuming the dark matter to be fermions, we must at a minimum  
introduce a new scalar or vector boson particle/field $B$, with mass $M_B$,
to mediate the dark matter annihilation to a neutrino-antineutrino pair.  
For the process $\chi + \bar\chi \rightarrow \nu \bar\nu$ 
(here we are using the generic $\nu$ to include all three neutrino flavors), one can imagine 
two types of vertices and diagrams:  either an s-channel tree graph (I) with vertices mediated by 
$\mathcal{L}_{{\chi\chi B}}$ and $\mathcal{L}_{\nu\bar{\nu}B}$, or a t-channel tree graph (II) with both 
vertices mediated by $\mathcal{L}_{{\chi B\nu}}$.
A new quantum number is needed to elevate the $\bar\chi\chi\bar\nu\nu$ operator to a privileged status,
but we do not indulge in detailed model building here.  Rather, we 
focus on the Standard Model (SM) loop-corrections to the diagrams (I) and (II).

\vspace{0.5cm}

\begin{center}
\begin{picture}(300,100)(0,0)
\ArrowLine(-20,100)(25,55)
\ArrowLine(25,55)(-20,10)
\DashLine(25,55)(80,55){2}
\ArrowLine(80,55)(125,100)
\ArrowLine(125,10)(80,55)
\Text(-20,77.5)[]{$\chi$}
\Text(-20,32.5)[]{$\overline{\chi}$}
\Text(52.5,65)[]{$B$}
\Text(120,77.5)[]{$\nu$}
\Text(120,32.5)[]{$\bar{\nu}$}
\ArrowLine(170,100)(242.5,100)
\ArrowLine(242.5,10)(170,10)
\ArrowLine(242.5,100)(315,100)
\ArrowLine(315,10)(242.5,10)
\DashLine(242.5,100)(242.5,10){2}
\Text(180,90)[]{$\chi$}
\Text(180,20)[]{$\overline{\chi}$}
\Text(252.5,55)[]{$B$}
\Text(305,90)[]{$\nu$}
\Text(305,20)[]{$\bar{\nu}$}
\Text(52.5,-20)[]{(I) s-channel B-exchange}
\Text(242.5,-20)[]{(II) t-channel B-exchange}
\end{picture}\\
\end{center}

\vspace{1.0cm}

There are order $(g^2)$ radiative corrections to the s-channel process which produce two-particle 
charged leptonic final states $\sum_l l^+l^-$.
Since we are comparing to the two-particle final states $\sum_l \nu_l{\bar\nu}_l$,
the number of flavor generations will cancel from our results.
For example, the ratio is unchanged whether the new interaction is assumed to produce all three neutrino flavors,  
with concomitant induced production $e^+e^-,\mu^+\mu^-,\tau^+\tau^-$,
or whether the new interaction is assumed to produce, say, just $\nu_\tau\bar\nu_\tau$ with 
concomitant production of just $\tau^+\tau^-$.
Corrections to the two-particle s-channel process are 
(a) s-channel Z-exchange, and (b) t-channel W-exchange.  
(These SM processes necessarily violate conservation of any charge carried by the $B$ particle.)

\vspace{0.5cm}

\begin{center}
\begin{picture}(300,100)(0,0)
\ArrowLine(-20,100)(2.5,55)
\ArrowLine(2.5,55)(-20,10)
\DashLine(2.5,55)(35.5,55){2}
\Photon(69.5,55)(102.5,55){4}{4}
\ArrowLine(102.5,55)(125,100)
\ArrowLine(125,10)(102.5,55)
\ArrowArc(52.5,55)(17,0,180)
\ArrowArc(52.5,55)(17,180,360)
\Text(0,92.5)[]{$\chi$}
\Text(0,17.5)[]{$\overline{\chi}$}
\Text(18.5,65)[]{$B$}
\Text(86,65)[]{$Z$}
\Text(52.5,80)[]{$\nu$}
\Text(52.5,30)[]{$\nu$}
\Text(110,92.5)[]{$l^-$}
\Text(110,17.5)[]{$l^+$}
\ArrowLine(170,100)(192.5,55)
\ArrowLine(192.5,55)(170,10)
\ArrowLine(242.5,55)(278.75,77.5)
\ArrowLine(278.75,32.5)(242.5,55)
\ArrowLine(278.75,77.5)(315,100)
\ArrowLine(315,10)(278.75,32.5)
\DashLine(192.5,55)(242.5,55){2}
\Photon(278.75,77.5)(278.75,32.5){4}{4}
\Text(190,92.5)[]{$\chi$}
\Text(190,17.5)[]{$\overline{\chi}$}
\Text(217.5,65)[]{$B$}
\Text(260.625,72.5)[]{$\nu$}
\Text(260.625,37.5)[]{$\bar{\nu}$}
\Text(290,92.5)[]{$l^-$}
\Text(290,17.5)[]{$l^+$}
\Text(290,55)[]{$W$}
\Text(52.5,-20)[]{(a) oblique correction}
\Text(242.5,-20)[]{(b) induced $B\,l^+ l^-$ vertex}
\end{picture}\\
\end{center}

\vspace{1.0cm}

The ``oblique'' correction in (a) induces $B$-$Z$ mixing.  One may choose a counterterm to 
eliminate the mixing at one scale $\mu$.  Although not natural, one can in principle choose the scale 
$\mu$ to eliminate the mixing at the threshold $\mu = 2M_{\chi}$; then, for non-relativistic DM, 
there is almost no production of $l^+l^-$ from non-relativistic $\chi$ annihilation.  
Similarly, the $Bl^+l^-$ vertex correction in 
(b) can be canceled by a counterterm at one scale, e.g. $\mu = 2M_{\chi}$.

Radiative corrections to the t-channel process which produce an $l^+l^-$ final state are an 
s-channel $Z$-exchange and a t-channel $W$-exchange.

\vspace{0.5cm}

\begin{center}
\begin{picture}(300,100)(0,0)
\ArrowLine(-20,100)(12.5,100)
\ArrowLine(12.5,10)(-20,10)
\DashLine(12.5,100)(12.5,10){2}
\ArrowLine(12.5,100)(52.5,55)
\ArrowLine(52.5,55)(12.5,10)
\ArrowLine(92.5,55)(125,100)
\ArrowLine(125,10)(92.5,55)
\Photon(52.5,55)(92.5,55){4}{4}
\Text(-10,90)[]{$\chi$}
\Text(-10,20)[]{$\overline{\chi}$}
\Text(5,55)[]{$B$}
\Text(27.5,92.5)[]{$\nu$}
\Text(27.5,17.5)[]{$\bar{\nu}$}
\Text(72.5,65)[]{$Z$}
\Text(110,92.5)[]{$l^-$}
\Text(110,17.5)[]{$l^+$}
\ArrowLine(170,100)(218,100)
\ArrowLine(218,100)(267,100)
\ArrowLine(267,100)(315,100)
\ArrowLine(218,10)(170,10)
\ArrowLine(267,10)(218,10)
\ArrowLine(315,10)(267,10)
\DashLine(218,100)(218,10){2}
\Photon(267,100)(267,10){4}{4}
\Text(280,55)[]{$W$}
\Text(200,55)[]{$B$}
\Text(194,90)[]{$\chi$}
\Text(194,20)[]{$\overline{\chi}$}
\Text(242.5,90)[]{$\nu$}
\Text(242.5,20)[]{$\bar{\nu}$}
\Text(291,90)[]{$l^{-}$}
\Text(291,20)[]{$l^{+}$}
\Text(52.5,-20)[]{(c) induced $Z\chi\bar\chi$ vertex}
\Text(242.5,-20)[]{(d) induced $\chi\bar\chi\nu\bar\nu$ box graph}
\end{picture}\\
\end{center}

\vspace{1.0cm}

In graph (c), the radiatively induced $\chi\bar\chi Z$ vertex can be canceled by a counterterm at one value of 
$\mu$, say $\mu = 2M_{\chi}$.  In contrast, the four-fermion vertex induced in graph (d) is of dimension 
six and has no counterterm.  Thus, it must be rendered finite by direct calculation.  
The superficial degree of divergence in graph 
(a) is quadratic, (b) is quadratic (in unitary gauge), (c) is logarithmic for scalar $B$ and quadratic 
for vector $B$ (in unitary gauge) and (d) is logarithmic for scalar B and quadratic for vector $B$ 
(in unitary gauge).  We now focus our attention on graph (d).
We do so for two reasons.  
The first reason is that there is no counterterm for this graph to suppress the operator,
and the second reason is that graph (d) is the least divergent of the graphs (along with (c)) 
and therefore represents 
a conservative window to our study of the stability of a ``neutrinos only'' model 
in the face of radiative corrections.

For our calculation, we make the simplest choice for the spin-parity of
the $B$-meson, namely that it is a scalar particle. 
In fact, we show in Appendix I that the divergent contribution
to the amplitude of the box graph factorizes into a divergent factor
independent of the
$B$-meson's spin-parity, times a factor proportional to the tree-level
amplitude.
This latter factor depends on the $B$-meson's spin-parity.
However, this latter factor cancels in the ratio of box to tree
amplitudes,
which is what we investigate.
Thus, our results are true for any spin-parity of the $B$-meson.

We render the box graph (d) finite by use of a Pauli-Villars (PV) regulator to compensate the $B$ particle propagator.  
The Pauli-Villars cutoff $\Lambda$ then becomes the maximum scale for a credible effective theory.  
We work in the unitary gauge, since this gauge, and only this gauge, includes only physical particles
throughout the calculation.  Thus, the only new parameter introduced is the PV regulator $\Lambda$,
which now takes on the significance as the mass-scale above which additional new particles enter.   
Some details of our calculation in the unitary gauge are presented in Appendix~(\ref{AppI}).
In Appendix~(\ref{AppII}) we calculate the same box graph in the general $R_\xi$-gauge, 
and compare the results to that from the unitary gauge calculation.
 
The amplitude for graph (d) depends on four dimensionful parameters, $M_\chi$, $M_B$, $M_W$, and $\Lambda$.  
Therefore, the ratio of rates (independent of the number of flavor generations)
\beq\label{rateratio1}
{\mathcal R}=\frac{\langle v\,\sigma(\chi\bar\chi\rightarrow l^+l^-)\rangle}
                  {\langle v\,\sigma(\chi\bar\chi\rightarrow \nu\ \bar\nu\ )\rangle}
\eeq
depends on three independent ratios, which we may take to be
$M_\chi/M_W$, $M_B/M_\chi$, and $\Lambda/M_B$.
$M_W$ is known, of course.  For the case of
a thermal relic (only), $M_\chi$ and $M_B$ are constrained by the requirement that the 
$\chi$ dark matter decouple from early universe thermal equilibrium such that $\Omega_{DM}h^2 = 0.1$,
where $h$ is the Hubble parameter in units of 100 km sec$^{-1}$ Mpc$^{-1}$.
We will investigate this constraint below, in Eqs.~(\ref{annrate}-\ref{massratiobnd}).

\begin{figure}[p]
\includegraphics[angle=270,width=.8\textwidth]{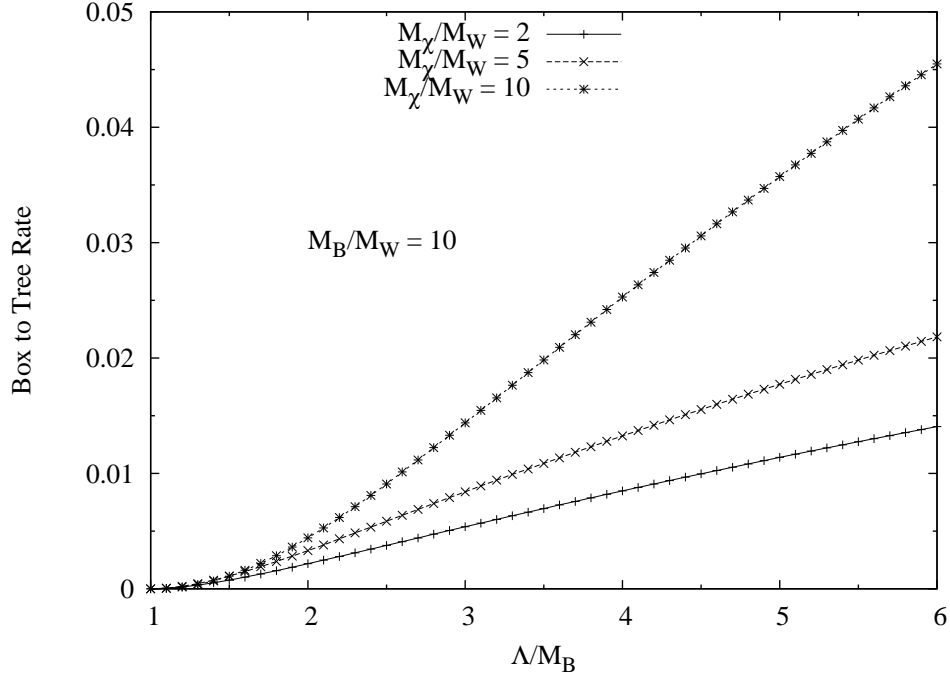}%
\caption{The ratio of the rate for $\chi \chi \rightarrow l^+ l^-$
from the box diagram to the rate for $\chi \chi \rightarrow \bar \nu \nu$
at tree level, as a function of $\Lambda/M_B$, where
$\Lambda$ is the cutoff scale of the effective theory and $M_B$ is
the mass of the boson that mediates the $\chi$ annihilation
into neutrinos.  Here, $M_B/M_W = 10$, and $M_\chi/M_W$ has the values 2, 5, and 10.}
\label{fig:Lambda1}
\end{figure}

\begin{figure}[p]
\includegraphics[angle=270,width=.8\textwidth]{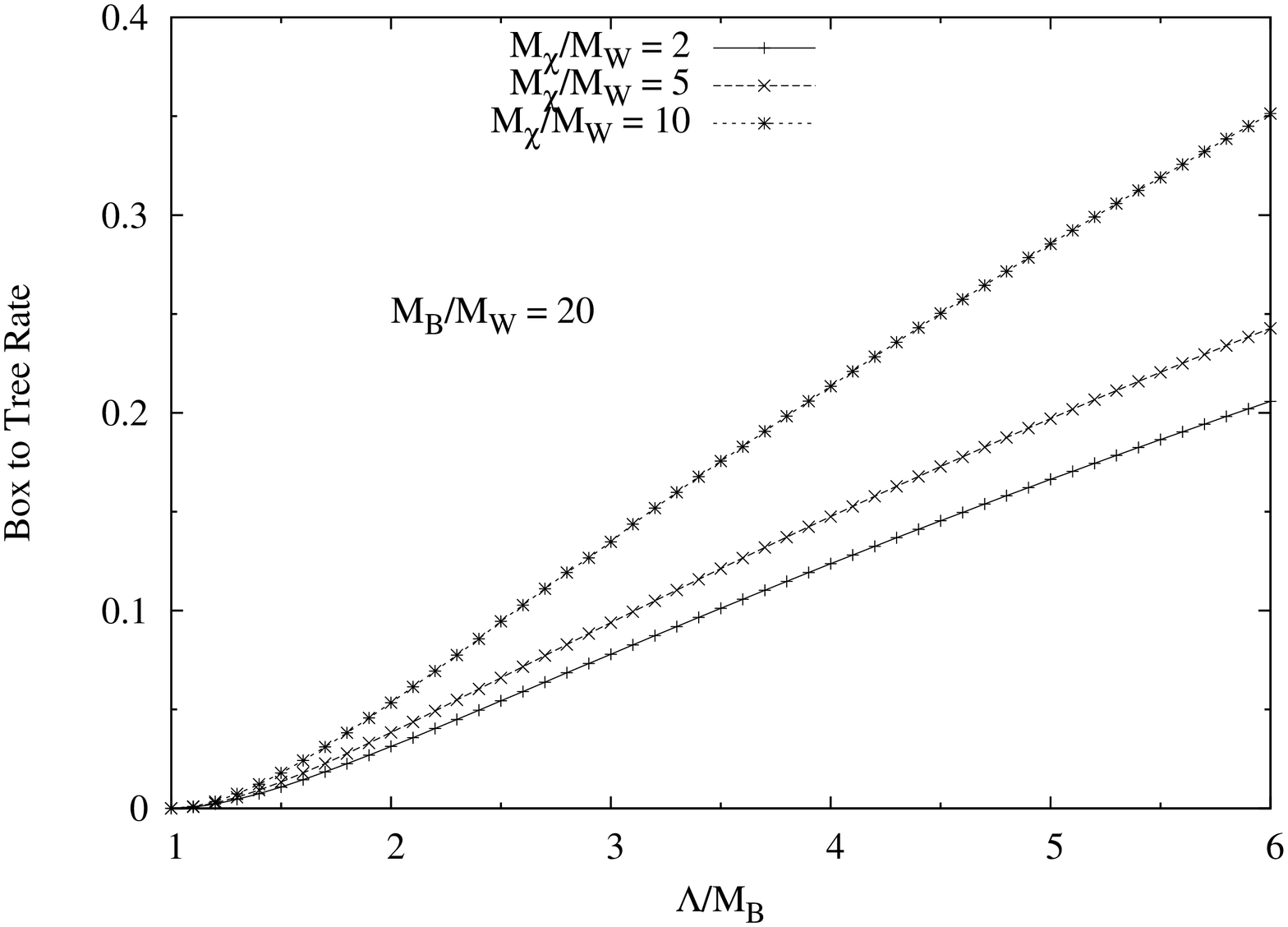}%
\caption{As Fig.~\ref{fig:Lambda1}, but with $M_B/M_W = 20$.}
\label{fig:Lambda2}
\end{figure}

The PV mass $\Lambda$ is a priori a free parameter.
Inherent in the PV regularization scheme is vanishing of the electromagnetic amplitude at $\Lambda=M_B$.
This results because the propagator for the fictitious unphysical 
$\Lambda$ ``particle'' enters with the wrong sign, by construction.
However, the spirit of the regularization is to produce a well-defined effective theory, 
valid up to the scale of the fictitious PV ``particle'' $\Lambda\gg M_B$.
The higher the value of $\Lambda$, the more stable is the effective theory.
We are therefore led to ask, at what value of $\Lambda/M_B$ does the effective theory for the 
``neutrinos only'' ansatz fail to support the idealized ``neutrinos only'' model?
We answer this question with the graphs displayed in Figs.~\ref{fig:Lambda1} and \ref{fig:Lambda2}.
If the theory fails at $\Lambda$ not much larger than $2M_\chi$, 
then the field theory is very incomplete as written; more fields are needed already near the dark 
matter scale $M_\chi$.  With more fields come more counterterms.  Consequently,
more fine tuning of these counterterms is needed to maintain the ``neutrinos
only" model.  It seems more likely that nature eschews this additional tedium,
and the electromagnetic branching ratio increases at scales $\agt \Lambda$.

In Figs.~\ref{fig:Lambda1} and \ref{fig:Lambda2} we show the ratio ${\cal R}$ 
of annihilation rates into $l^+l^-$ (through the box diagram) versus into $\nu\bar\nu$,
as a function of the cutoff scale (in units of $M_B$), 
with $M_B$ taken to be 10 and 20 times the $W$ mass,
respectively,  
and for $M_\chi$ taken to be 2, 5, and 10 times $M_W$.
In calculating the box graph, we have neglected complicating non-leading contributions,
as they are unnecessary for the thesis of this paper.
From the figures, one sees that already at $\Lambda\sim 3$~times $M_B$, the relative rate into the $l^+l^-$
mode is of order $1\%$.
In fact, we may invoke the simple asymptotic formula in Eq.~(\ref{rateratio3}) to infer that 
the rate ratio ${\cal R}$ grows as $\ln^2 (\Lambda/M_B)$, and that a somewhat 
minimal mass ordering $\Lambda\sim 10\,M_B\sim 10^2\,M_W$ will necessarily yield  
${\cal R}\sim$~few percent. 

%
\begin{figure}[p]
\includegraphics[angle=270,width=.8\textwidth]{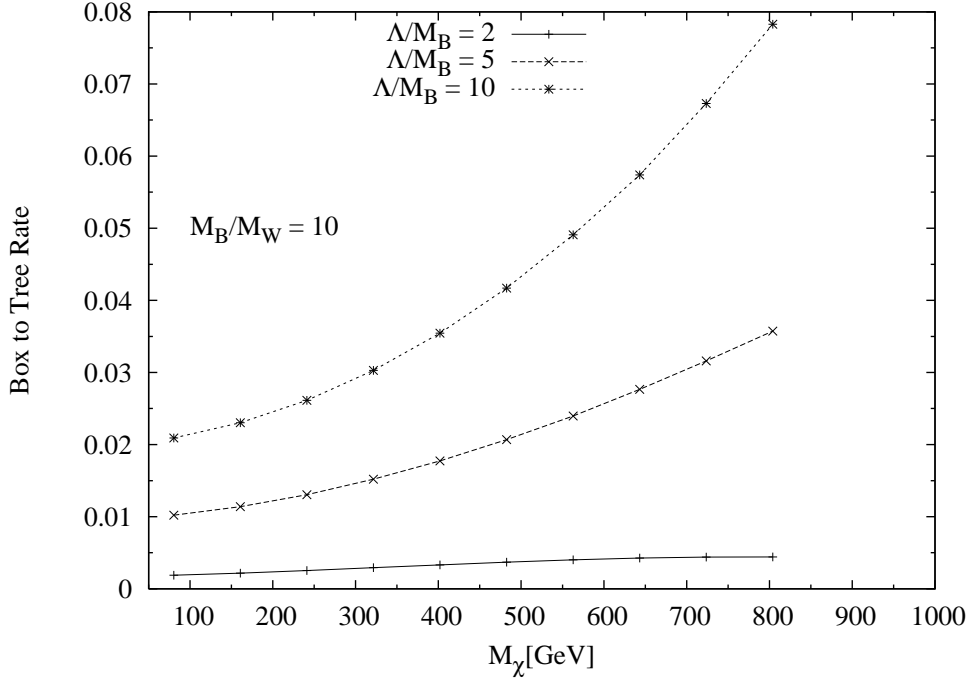}
\caption{The ratio of the rate for $\chi \chi \rightarrow l^+ l^-$
from the box diagram to the rate for $\chi \chi \rightarrow \bar \nu \nu$
at tree level, as a function of the $\chi$ mass, for
$M_B/M_W = 10$ and $\Lambda/M_B=2$,~5,~and~10.
$M_B$ is the mass of the boson that mediates the $\chi$ annihilation
into neutrinos, and $\Lambda$ is the cut-off scale of the effective theory.}
\label{fig:Rate1}
\end{figure}
\begin{figure}[p]
\includegraphics[angle=270,width=.8\textwidth]{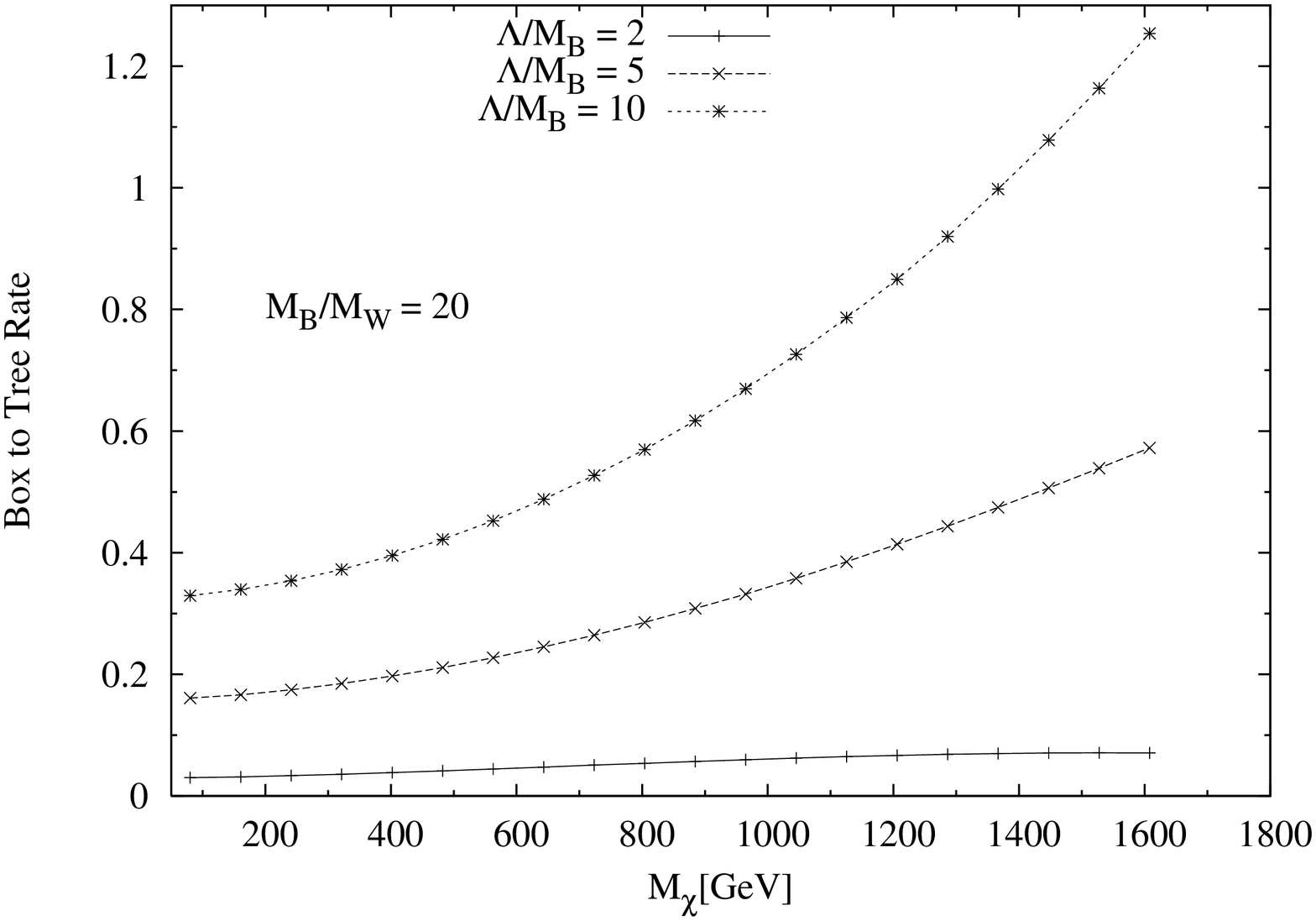}%
\caption{As Fig.~\ref{fig:Rate1}, but with $M_B/M_W = 20$.}
\label{fig:Rate2}
\end{figure}
%

Our results for the ratio ${\cal R}$ of rates
versus the $\chi$ mass are shown in Figs.~\ref{fig:Rate1} and \ref{fig:Rate2}.
We note that ${\mathcal R}$ is an increasing function of $M_\chi$,
of $\Lambda/M_B$, and of $M_B/M_W$, as shown explicitly in Eq.~(\ref{rateratio2}).  
We note that the monotonic increase of the rate ratio with $\Lambda$ is an inevitable 
consequence of PV regularization, and the increase with $M_\chi$ and $M_B$ derives from 
the propagator suppression of the tree-level rate, 
$\sim(t-M_B^2)^{-2}\approx (-M_\chi^2-M_B^2)^{-2}$. 
Fortunately, it is precisely the most physically-reasonable
values for these parameters that yield the largest electromagnetic branching ratio.
For a reasonable renormalization scheme, we would want $\Lambda \gg M_B$.
Further, $M_B/M_W$ must be large enough to avoid heretofore undetected new physics~\cite{bilenky}.

It is also interesting to determine the constraints which apply to our parameters
if $\chi$ is a thermal relic.
In this case, $M_B$ and $M_\chi$ are constrained by the
requirement that the $\chi$
abundance account for the currently observed dark matter:
the value $\Omega_{\chi}h^2 = 0.1$ requires an annihilation rate of
approximately $\langle v \sigma \rangle \sim 3 \times 10^{-26}$ cm$^3$ sec$^{-1}$ for 
thermally-produced dark matter~\cite{scherrer,jungman}.
In the model presented here,
in the non-relativistic limit $s\sim (2\,M_\chi)^2$ and large $M^2_B\gg s$ limit, we have
\beq
\label{annrate}
\langle v\,\sigma\rangle \times Br(\chi\bar\chi\rightarrow\nu\bar\nu)
= \left(\frac{g_B^4}{4\pi}\right) \left(\frac{M^2_\chi}{M^4_B}\right)
\eeq
for the pure $L=0$ partial wave $s$-channel exchange of the $B$, 
and $\sim\half$ times this for $t$-channel $B$ exchange 
(where an $L=1$ contribution to the rate is suppressed by the square of the $\chi$~velocity);
$Br(\chi\bar\chi\rightarrow\nu\bar\nu)$ is the annihilation branching ratio to $\nu\bar\nu$. 
Substituting the desired rate for $\langle v\,\sigma\rangle$ into this equation,
and taking $Br(\chi\bar\chi\rightarrow\nu\bar\nu)=1$,
we obtain the result
\begin{equation}
\label{constrain1}
\left(\frac{4\pi}{g_B^2}\right)
\left(\frac{M_B}{M_\chi}\right)^2 M_\chi = 70~{\rm TeV},
\end{equation}
or
\begin{equation}
\label{constrain2}
\frac{M_B}{M_\chi}=26\,\sqrt{\left(\frac{g_B^2}{4\pi}\right)\left(\frac{100~{\rm GeV}}{M_\chi}\right)}.
\end{equation}
If we consider only $M_\chi > 100$ GeV, and restrict the coupling to the perturbative regime
$g_B^2 \le 4\pi$, then we obtain the bound~
\footnote
{
We note that a weaker but more rigorous bound comes from unitarity of the annihilation process.
The textbook unitarity bound is simply 
$\sigma\le 4\pi\,(2L+1)/p_{\rm cm}^2$ for the $l^{th}$ partial wave,
from which we infer 
$\langle v\,\sigma\rangle\le 4\pi/(v\,M_\chi^2)$
for non-relativistic particles in the $L=0$ partial wave.
Substituting for the LHS from Eq.~(\ref{annrate}),
one obtains $(g_B^2/4\pi)\le \frac{1}{\sqrt{v}}\,(M_B^2/M_\chi^2)$.
Thus, unitarity allows strong coupling, 
if it is suppressed in the amplitude by the propagator factor $\sim M_B^{-2}$.
The bound from the perturbative assumption is not so forgiving.
}
\begin{equation}
\label{massratiobnd}
1 < \frac{M_B}{M_\chi} < 26,
\end{equation}
where the lower bound comes from the additional requirement that 
$M_\chi < M_B$ so that the $\chi$ dark matter is not de-stabilized by the 
simple decay chain $\chi\rightarrow B+\nu$~
\footnote
{
For $M_B<M_\chi$, one finds the decay width 
$\Gamma(\chi\rightarrow B+\nu)=\frac{g_B^2}{16\pi}\left(1-\frac{M_B^2}{M_\chi^2}\right)^2 M_\chi$.
}.
Thus,
the mass of the boson $B$ cannot be wildly larger than that
of the dark matter particle.  
(When the box-diagram annihilation rate
into $l^+l^-$ becomes significant, 
the value of $Br(\chi\bar\chi\rightarrow\nu\bar\nu)<1$
must be included in the algebra.)  Of course, if the dark matter is
not a thermal relic, then these constraints no longer apply.  None
of the other results in this paper rely on the assumption
that $\chi$ is a thermal relic, and of course, the constraints on
$\langle v \sigma \rangle$ that we examine later in the paper
are irrelevant if $\langle v \sigma \rangle$ is fixed
to give the thermal relic abundance.

We have presented figures only for the case where $M_\chi > M_W$.
However, the calculation in Appendix~(\ref{AppI}) is valid for the case $M_\chi < M_W$ as well.  
In this case, we find that the branching into $l^+ l^-$ is infinitesimal.
However, as we have not included the induced $Z\chi\bar\chi$ vertex shown in 
Fig.~(c) in the calculation of the annihilation rate into $\nu \bar \nu$,
we cannot offer a reliable estimate of the rate if $M_\chi < M_W$.
A calculation of the contribution of the vertex diagram to the total annihilation
rate is irrelevant to conclusions of the present paper, but would be needed
to extend the framework presented here
to a more complete ``neutrinos only" annihilation model for $M_\chi < M_W$.

Returning to the case $M_\chi > M_W$, 
our results indicate that a model for dark matter
annihilation that produces only neutrinos at tree level, 
when sensibly embedded into a field theory,
will produce a significant ($> 1\%$) branching into $l^+l^-$.
It is reasonable to ask whether such a branching ratio is large
enough to provide limits competitive with the neutrino limits
discussed in Refs.~\cite{beacom,yuksel,mack}.
The neutrino annihilation limits
in Ref.~\cite{beacom} are for cosmic dark matter annihilations, while
the (tighter) limits from Ref.~\cite{yuksel} are for annihilations in the halo
of the galaxy; it is the latter which we will consider here.  (Palomares-Ruiz
and Pascoli~\cite{PR} consider neutrino constraints for
dark matter masses below 100 MeV, 
outside of the mass range for which our results are interesting).
Ref.~\cite{mack} also provides general limits on annihilation into high-energy
photons.  These limits are not directly applicable to our calculation,
since we consider branching into charged leptons rather than photons.
High-energy leptons will convert their energy into photons
primarily through inverse Compton scattering off of background photons,
or, if a magnetic field is present, through synchrotron radiation.
For our purposes, it is sufficient to consider only the latter effect,
within the Galaxy, where the magnetic field is reasonably well known.
For discussions of the former effect,
see, e.g., Refs.~\cite{BW,Cola1,Cola2,Profumo,Jeltema}.
(See also Ref. \cite{Zel} for one of the earliest discussions of
these phenomena).

Synchroton radiation from the products of
dark matter annihilation has been investigated in some detail
\cite{BW,Cola1,Cola2,B1,B2,Gondolo,Bertone,BOT,ABO}.
More recently, it has been suggested that
the microwave ``haze" near the center of the Galaxy, observed by WMAP,
could be synchrotron radiation from such
annihilations~\cite{Finkbeiner,Hooper1,Hooper2}.  Finally,
Hooper~\cite{Hooper3} has used the WMAP observations to bound
annihilation-produced synchrotron radiation.  This latter
result can be combined with our predicted branching ratio
into charged leptons and compared with the corresponding neutrino bounds.

Our results allow branching
into $e^+e^-$, $\mu^+\mu^-$, or $\tau^+\tau^-$, depending on
the neutrinos that couple in ${\cal L}_{\chi B\nu}$.  
Synchrotron radiation is produced directly by $e^+e^-$, and indirectly through
the decay-produced electrons from $\mu^+\mu^-$ and $\tau^+\tau^-$.
In the latter two cases, some of the decay energy goes into neutrinos and does
not contribute to the synchrotron radiation.  Thus, the $e^+e^-$ annihilation
channel yields the strongest synchrotron signal, followed by $\mu^+\mu^-$ and
then $\tau^+\tau^-$~\cite{Hooper1}.  Since we make no prediction for
the specific type of neutrinos produced by the dark matter annihilations,
we can derive the most conservative limits by considering annihilation
into $\tau^+\tau^-$ only.

Hooper~\cite{Hooper3} gives limits on
the cross-section for dark matter annihilation into $\tau^+\tau^-$ using
the requirement that the resulting synchrotron radiation not exceed
the WMAP microwave observations near the center of the Galaxy.
He presents two sets of limits, the first corresponding to
an assumed Navarro-Frenk-White (NFW) profile \cite{NFW}, and the second
to an assumed uniform distribution of the dark matter inside of the solar
circle.  Similarly, Yuksel et al. \cite{yuksel} give three different
sets of limits based on the angular scale over which the neutrino
emission is measured, and they provide appropriate modification factors
for each limit assuming any of three density profiles.
While there is no exact correspondence between any of the limits provided
in Ref. \cite{Hooper3} and those given in Ref. \cite{yuksel},
the closest correspondence is between the NFW sychrotron limit
in Ref. \cite{Hooper3} and the ``Halo Angular" limit in Ref. \cite{yuksel},
for the parameters appropriate to the NFW profile.  Comparing these
two sets of limits for
annihilation purely into 
$\tau^+\tau^-$, with $M_\chi = 100$ GeV, we find that the upper bound on
$\langle v \sigma \rangle$ from $\bar \nu \nu$
production is roughly
$1000$ times the corresponding limit from $\tau$-produced
synchrotron radiation.
This ratio decreases to about 30
at $M_\chi = 1$ TeV.  Thus, the synchrotron bounds from the production
of charged leptons are tighter than the neutrino bounds
at $M_\chi\sim 100$~GeV if the branching ratio to
charged leptons is $\agt 0.1\%$.  For $M_\chi\sim$~1 TeV, this
branching ratio must be $\agt 3\%$ in order for the synchrotron bounds
to be tighter than the neutrino bounds.
This is the mass range (100~GeV~$\alt M_\chi \alt$ 1~TeV) of 
greatest interest from the point of view of WIMP dark matter. 
(Strictly speaking, the neutrinos from $\tau$ decay should be included
in the total neutrino signal when comparing to the neutrino
limits in Ref.~\cite{yuksel}, but this is a small
effect as long as the branching ratio into charged leptons
is $\ll 1$).  

Our results
do not undercut the neutrino limits proposed in Refs.~\cite{beacom,yuksel,PR};
these certainly remain valid limits on the annihilation rate for
dark matter.  Further, it is clear that both the synchrotron limits
derived in Ref.~\cite{Hooper3} and the neutrino limits
in Ref. \cite{yuksel} depend on
the assumed model for the distribution
of dark matter in the Galaxy.  However, our purpose in this
paper is not to determine the exact ratio between these two sets
of limits (which may change with improved observations or 
more detailed calculations
in any case) but simply to point out that
the neutrino and charged lepton/photon sensitivities 
are likely to be competitive with each other over the WIMP
mass range of greatest interest.
For dark matter masses below the $W$ mass, we can offer
no useful new limits.  Our calculations in this case allow annihilation into
neutrinos only, with only an infinitesimal branching to other particles.

Although our results might seem to be limited to the specific model
we have examined, they are rather general.  Whenever
neutrino-antineutrino pairs are produced in the final state,
they can always be converted into $l^+l^-$
pairs through $Z$ or $W$ exchange as in diagrams (a) to (d) above.
This is an inevitable consequence of the Standard Model; 
no new physics is required.
The only assumption concerns the creation of a specific interaction 
that couples dark matter to $\nu \bar \nu$ pairs at tree level.  
Once such a model is in place, however, 
the induced coupling to $l^+l^-$ is inevitable and readily calculable.

\acknowledgments

J.B.D., R.J.S., and T.J.W. were supported in part by the Department of Energy (DE-FG05-85ER40226).
We thank J.\ Beacom, D.\ Hooper, and G.\ Mack for helpful discussions.

\section{Appendix: Calculation of Box Diagram (in Unitary Gauge) and Electromagnetic Rate}
\label{AppI}

Here we present the amplitude for the box diagram of Figure (d),
and a calculation of the divergent part of this amplitude.  
Then we construct the ratio of rates, 
\beq
{\mathcal R}=\left[
\frac{\langle v\,\sigma(\chi\bar\chi\rightarrow l^+l^-)\rangle}
                  {\langle v\,\sigma(\chi\bar\chi\rightarrow \nu\ \bar\nu\ )\rangle}
\right]_{DivPart}\,.
\eeq
The subscripted qualifier ``DivPart'' is a reminder that only the leading (divergent) contribution 
to the electromagnetic rate will be included here.

In unitary gauge, the $W$-propagator is 
\beq\label{Wpropagator1}
\frac{-i\,(g_{\mu\nu}-k_{\mu}k_{\nu}/M_{W}^2)}{k^2-M_{W}^2+iM_{W}\Gamma_{W}}\,,
\eeq
and the matrix element for the amplitude of the box (d) is given by
\begin{eqnarray}\nonumber
i \mathcal{M}_{\rm box} &=&\int\frac{d^4 k}{(2\pi)^4}\left(\frac{i}{(k-p+q)^2-M_{B}^2 
  + iM_B\Gamma_B}\right)\left(\frac{-i\,(g_{\mu\nu}-k_{\mu}k_{\nu}/M_{W}^2)}{k^2-M_{W}^2+iM_{W}\Gamma_{W}}\right)\\
&&\hspace{1.0cm}
     \times\,\left(\frac{i}{(k+q)^2 -M_{\nu}^2 + i\epsilon}\right)\left(\frac{i}{(\bar{q}-k)^2 -M_{\bar{\nu}}^2 + i\epsilon}\right)\\\nonumber
&& \\\nonumber
&&\times[\bar{u}(q)\Gamma_{W}^{\mu}(\slashed{k} + \slashed{q} + M_{\nu})\Gamma_B u(p)]\,
	[\bar{v}(\bar{p})\Gamma_B(\slashed{\bar{q}} - \slashed{k} + M_{\bar{\nu}})\Gamma_{W}^{\nu}v(\bar{q})]\,,
\end{eqnarray}
where the momenta assignments to the particles are $\chi(p)$, $\overline{\chi}(\bar{p})$, $l^{-}(q)$, $l^{+}(\bar{q})$, $W(k)$,
which in turn determine the further assignments $B(k-p+q)$, $\nu(k+q)$, and $\bar{\nu}(\bar{q}-k)$. 
Here $\Gamma^{\mu}_W=(\frac{g}{\sqrt{2}})\gamma^\mu \frac{(1-\gamma_5)}{2}$ is the usual charged-current electroweak vertex  
with the $SU(2)$~coupling constant $g=e/\sin\theta_w$,
$p$ is the four-momentum of the dark matter, 
$q$ is the four-momentum of the electron, and the internal loop four-momentum of the virtual $W$-boson is given by $k$.  
In addition, $\Gamma_B$ in the numerator is the coupling times Lorentz structure assigned to a $B\chi\nu$ vertex,
not to be identified with the $B$ width in the denominator.
The value of the coupling we do not require, 
as the coupling cancels out when divided by the amplitude of the tree diagram;
accordingly, we let simplicity be our guide and choose a scalar Lorentz structure for the $B$~field. 
We will neglect the neutrino and charged lepton masses, and define $M_\chi$ to be the dark matter mass.

We introduce four Feynman parameters $\xi_i$ (one per internal line) 
and find for the denominator $D$:
\begin{eqnarray}
\frac{1}{D} &=& (4-1)!\int_{0}^{1}d\xi_1\int_{0}^{1}d\xi_2\int_{0}^{1}d\xi_3\int_{0}^{1}d\xi_4\,
   \delta(1-[\xi_1+\xi_2+\xi_3+\xi_4])\\\nonumber
&&\times\,\left[\xi_1((k-p+q)^2-M_{B}^2)+\xi_2(k^2-M_{W}^2) + \xi_3(k+q)^2 + \xi_4(k-\bar{q})^2\right]^{-4}
\end{eqnarray}
Here we have dropped the $B$ and $W$ widths from the denominator, as they play no important role in the 
present calculation. 
Upon rotating $k$ to Euclidean space, we find that the divergent part of the amplitude,
coming from the longitudinal mode of the $W$~propagator, is
\beq
\mathcal{M}_{\rm box}=\frac{-g^2}{8 M_{W}^2}\int \frac{d^4k_E}{(2\pi)^4}\int d\xi\frac{k_{E}^4}{(k_{E}^2 + \Delta_B^2)^4}
  \times[\bar{u}(q)(1 + \gamma_5)\Gamma_Bu(p)]\,[\bar{v}(\bar{p})\Gamma_B(1-\gamma_5)v(\bar{q})]\,.
\eeq
%
Here we have defined the functional 
\begin{eqnarray}
\int d\xi &\equiv& 
   3!\int_{0}^{1}d\xi_1\int_{0}^{1}d\xi_2\int_{0}^{1}d\xi_3\int_{0}^{1}d\xi_4\,
   \delta\left(1-[\xi_1+\xi_2+\xi_3+\xi_4]\right)\\\nonumber
  &=&
   6\left[ \int_{0}^{1}d\xi_1\int_{0}^{1-\xi_1}d\xi_2\int_{0}^{1-\xi_1-\xi_2}d\xi_3
    \right]_{\xi_4=1-\xi_1-\xi_2-\xi_3}
\end{eqnarray}
and the parametrized mass-squared 

\beq\label{DB}
\Delta_{B}^2 \equiv \xi_1\,M_{B}^2 + \xi_2\,M_{W}^2 
   -(\xi_1-\xi_{1}^2 - 2\xi_1\xi_2 +4\xi_{3}\xi_4)\,M^2_\chi\,.
\eeq
%
In (\ref{DB}) we have set the invariant $t\equiv (q-p)^2$ equal to the value appropriate
for a non-relativistic $\chi\bar\chi$~annihilation, namely, $t\approx -M_\chi^2$.

Performing the integral over $k_E$ will result in a logarithmic divergence, and therefore we introduce 
a Pauli-Villars regulator for the $B$ propagator
%
\beq
\frac{i}{(k-p+q)^2-M_{B}^2} \rightarrow \frac{i}{(k-p+q)^2-M_{B}^2} - \frac{i}{(k-p+q)^2-\Lambda^2}
\eeq
%
As is well-known, the Pauli-Villars regularization preserves local and global symmetries of the interaction.
We then perform the integral over $k_E$, and find that the divergent amplitude after regularization becomes
\begin{eqnarray}\label{Mbox}
\mathcal{M}_{\rm box}=\frac{-g^2}{2^7\pi^2 M_{W}^2}[\bar{u}(q)(1 + \gamma_5)\Gamma_B u(p)]\,
	[\bar{v}(\bar{p})\Gamma_B (1-\gamma_5)v(\bar{q})]\int d\xi\,\ln\left|\frac{\Delta_{\Lambda}^2}{\Delta_{B}^2}\right| \,,
\end{eqnarray}
where
%
\beq
\Delta_{\Lambda}^2 = \Delta_B^2 (M_B^2\rightarrow\Lambda^2)
   = \xi_1\,\Lambda^2 + \xi_2\,M_{W}^2 
   -(\xi_1-\xi_{1}^2 - 2\xi_1\xi_2 +4\xi_{3}\xi_4)\,M^2_\chi\,.
\eeq
%
Due to the assumed scalar nature of the $B$ particle, the $\gamma_5$'s in Eq.~(\ref{Mbox})
may be omitted.

The tree-level amplitude for $\chi\bar\chi\rightarrow\nu\bar\nu$ is 
\beq\label{Mtree}
i {\mathcal M}_{\rm tree}= [\bar{u}(q)\Gamma_B u(p)]\left(\frac{i}{t-M_B^2}\right)[\bar{v}(\bar{p})\Gamma_B v(\bar{q})]\,.
\eeq
Dividing the box amplitude by the tree-level amplitude (with $t\approx -M_\chi^2$, again) and squaring, 
we arrive at the result 
\beq\label{rateratio2}
{\mathcal R} = 
    \left(\frac{M_{B}^2+M_\chi^2}{128\cdot 8\pi\cdot M_W^2}\right)^2 
    \left[\ \int d\xi \ln\left|\frac{\Delta_{\Lambda}^2}{\Delta_{B}^2}\right|\ \right]^2\,.
\eeq
For the running $SU(2)$ coupling we use the value applicable at the weak scale: 
\beq
g^2=\frac{e^2}{\sin^2\theta_w}\approx 4\,e^2 =16\pi\,\alpha \approx \frac{16\pi}{128}\,.
\eeq
%
Notice that the spin-parity assignment of the $B$-meson appears only via
the
Dirac structure $\Gamma_B$, which cancels out of the ratio ${\cal R}$.
Thus, our results to follow are independent of the $B$-meson's spin and
parity.

There are some interesting, exactly calculable limiting cases.
For $M_B^2\gg M_\chi^2,\ M_W^2$, one has $\Delta_\Lambda^2/\Delta_B^2\approx \Lambda^2/M_B^2$,
independent of $\int d\xi$, and so Eq.~(\ref{rateratio2}) becomes to lowest non-vanishing order, 
simply
\beq\label{rateratio3}
{\mathcal R} \longrightarrow 0.97\times 10^{-7}\,
 \left(\frac{M_B}{M_W}\right)^4  
\left[\ \ln\left(\frac{\Lambda^2}{M_{B}^2}\right)\ \right]^2\,,
\quad {\rm for\ }M_B^2\gg M_\chi^2,\ M_W^2\,.
\eeq
We note that Eq.~(\ref{massratiobnd}) allows this mass ordering, but does not require it.

Another interesting calculable limit is 
$M_\chi^2\ll M_B^2, \Lambda^2, M_W^2$.  
Neglecting $M_\chi^2$ in $\Delta^2_B$ and $\Delta^2_\Lambda$, one finds that to lowest non-vanishing order,
\beq\label{rateratio4}
{\mathcal R} \longrightarrow  \left(\frac{M_{B}^2}{128\cdot 8\pi\cdot M_W^2}\right)^2 
 \left[\,
    \ln\left(\frac{\Lambda^2}{M_B^2}\right) 
    +\frac{\ln\left(\frac{M_W^2}{\Lambda^2}\right)}{\left(1-\frac{\Lambda^2}{M_W^2}\right)}
    -\frac{\ln\left(\frac{M_W^2}{M_B^2}\right)}{\left(1-\frac{M_B^2}{M_W^2}\right)} 
 \,\right]^2\,,
\ {\rm for\ }M_\chi^2\ll M_B^2, \Lambda^2, M_W^2\,.
\eeq
If in addition to $M_\chi^2\ll M_B^2,\Lambda^2, M_W^2$, one includes $M_B^2,\Lambda^2 \ll M_W^2$, 
i.e., a low-mass model with a low-mass cutoff,
then there results 
\beq\label{rateratio5}
{\mathcal R} \rightarrow  0.97\times 10^{-7} \left(\frac{M_B}{M_W}\right)^4
 \left[\,
    \left(\frac{\Lambda^2}{M_W^2}\right)\ln\left(\frac{M_W^2}{\Lambda^2}\right) 
   -\left(\frac{M_B^2}{M_W^2}\right)\ln\left(\frac{M_W^2}{M_B^2}\right)
 \,\right]^2,
\ {\rm for\ }M_\chi^2\ll M_B^2,\Lambda^2\ll M_W^2\,.
\eeq
Another limit, for a low-mass model with a high-mass cutoff, is
\beq\label{rateratio6}
{\mathcal R} \rightarrow  0.97\times 10^{-7} \left(\frac{M_\chi^2+M_B^2}{M_W^2}\right)^2
 \left[\,
    \frac{ \ln\left(\frac{\Lambda^2}{M_W^2}\right) }{ \left(1-\frac{M_W^2}{\Lambda^2}\right)}
 \,\right]^2,
\quad\quad\quad\ {\rm valid\ for\ }M_\chi^2, M_B^2\ll \Lambda^2, M_W^2\,.
\eeq
The bracketed quantity in (\ref{rateratio6}) is a monotonically increasing function of $\Lambda/M_W$, 
equal to zero at $\Lambda=0$ (as it must at $\Lambda=M_B$), to 1 at $\Lambda=M_W$,
and growing as $\ln(\Lambda^2/M_W^2)$ at $\Lambda^2\gg M_W^2$.
It is clear that with the ordering  $M_B^2\ll M_W^2$, 
i.e., when the tree graph has a light-mass ($M_B$) propagator while the box graph has a heavy-mass ($M_W$) 
propagator, then the resulting electromagnetic branching fraction is negligible for any value of the 
effective field thoery cutoff $\Lambda$, even up to the Planck mass.

\section{Appendix: Calculation of Box Diagram in {\boldmath $R_\xi$} Gauge}
\label{AppII}
Given that the unitary gauge admits only the three physical $W$ spin states into the $W$ propagator, 
this gauge choice is guaranteed to produce diagrams whose cuttings respect the Cutkowsky rules; 
other gauges will not.
Nevertheless, the ``renormalizable $R_\xi$ gauges'', rather than the unitary gauge, 
are often invoked because their resulting propagators 
offer a more benign high-energy behavior. Of course, when a gauge invariant set of graphs is summed to 
calculate an $S$-matrix element, all gauges must give the same gauge-invariant ($\xi$-independent) answer.
However, the box graph of interest here is not itself gauge invariant, the reason being that the 
dark sector breaks electroweak gauge invariance by treating the neutrino differently from its 
$SU(2)$ charged lepton partner.  Consequently, we expect the box graph to yield a gauge-dependent answer.
We have dealt with this situation in the main text 
by arguing that the unitary gauge is singled out since it, and it alone,
includes only physical $W$ states in its Feynman rules.
Nevertheless, it is interesting and instructive to calculate the box graph in a general $R_\xi$ gauge.
We do that here.

The unitary gauge $W$-propagator, given in Eq.~(\ref{Wpropagator1}), is the $\xi\rightarrow\infty$ limit of 
the general $R_\xi$-gauge $W$-propagator (we neglect the finite $W$-width) 
\beq\label{Wpropagator2}
D^W_{\mu\nu}(\xi)=\left( \frac{-i\,}{k^2-M_{W}^2} \right) 
   \left(g_{\mu\nu}-\frac{k_{\mu}k_{\nu}\,(1-\xi)}{k^2-\xi M_{W}^2}\right)\,,
\eeq
and the accompaning unphysical Goldstone boson propagator is 
\beq\label{GBpropagator}
D_{GB}(\xi)=\frac{+i}{k^2-\xi M_W^2}\,.
\eeq
In contrast to the calculation in the unitary gauge where the leading high-energy behavior was given by a 
logarithmic divergence, here, for any finite $\xi$, the leading high energy contribution to the box graph is finite 
and therefore calculable.  However, the result will be gauge ($\xi$) dependent.
To facilitate the calculation, we note that (i) the Goldstone boson contribution to the amplitude
does not contribute to the leading high-energy behavior, and so will be neglected,
and (ii) the $R_\xi$ gauge $W$-propagator (Eq.~\ref{Wpropagator2})
can be rewritten as a sum of the $\xi$-independent 
unitary gauge propagator and a $\xi$-dependent correction:
\beq\label{Wpropagator3}
D^W_{\mu\nu}(\xi)= \frac{-i\,(g_{\mu\nu}-k_{\mu}k_{\nu}/M_{W}^2)}{k^2-M_{W}^2}
          + \frac{+i\,(-k_\mu k_\nu/M_W^2)}{k^2-\xi M_W^2}\,.
\eeq
As far as the high-energy behavior of the amplitude is concerned, 
the contribution from the unitary gauge $W$-propagator again yields $\Delta^2_B$ in Eq.~(\ref{DB}), 
which we rename $\Delta^2_W$ to emphasize the role of the physical $W$-propagator in this contribution.
The additional contribution from the ``wrong sign'' propagator with a pole at the 
Goldstone boson mass-squared $\xi M_W^2$ yields
\beq\label{DGB}
\Delta_{GB}^2 \equiv \xi_1\,M_{B}^2 + \xi_2\,(\xi M_{W}^2) 
   -(\xi_1-\xi_{1}^2 - 2\xi_1\xi_2 +4\xi_{3}\xi_4)\,M^2_\chi\,.
\eeq
We then have 
\beq\label{rateratio-xi}
{\mathcal R} = 
    \left(\frac{M_{B}^2+M_\chi^2}{128\cdot 8\pi\cdot M_W^2}\right)^2 
    \left[\ \int d\xi \ln\left|\frac{\Delta_{GB}^2}{\Delta_{W}^2}\right|\ \right]^2\,.
\eeq
The bracketed expression here, 
$\int d\xi \ln\left|\frac{|\Delta^2_{GB}}{\Delta^2_W}\right| $,
is bounded by $\ln\left|\frac{\xi m_W^2}{M_W^2}\right| = \ln |\xi| $.  
We see that the fictitious Goldstone boson mass-squared $\xi M_W^2$ has come to play the role in the $R_\xi$-gauge calculation 
that the fictitious PV mass-squared $\Lambda^2$ played in the unitary gauge calculation.
As a consistency check, we see that both $\xi\rightarrow\infty$ and 
$\Lambda^2\rightarrow\infty$ return the infinite unitary gauge contribution without 
PV regularization.

This result explicitly demonstrates the merit of using the unitary gauge and a Pauli-Villars 
regularization scheme for the divergent graph.  
The particle degrees of freedom in the unitary gauge are all physical, 
and the PV cutoff $\Lambda$ has physical meaning.
$\Lambda$ is the scale up to which the model is applicable, and beyond which 
the ultraviolet completion of the theory cannot be ignored.
In contrast, the cutoff provided by the Goldstone mass 
in the renormalizable gauges has no physical interpretation.

\end{document}